\documentclass[10pt,twocolumn,twoside]{ieeeconf}
\IEEEoverridecommandlockouts    
\newtheorem{remark}{Remark}

\newtheorem{lemma}{Lemma}
\newtheorem{proposition}{Proposition}

\newtheorem{theorem}{Theorem}
\newtheorem{assumption}{Assumption}

\usepackage{cite}
\usepackage{amsmath}
\usepackage{amstext}
\usepackage{amssymb}
\usepackage{tikz}
\usepackage[mathscr]{euscript}

\usepackage{amsfonts}
\usepackage{enumerate}
\usepackage[final]{pdfpages}
\usepackage{csquotes}

\let\labelindent\relax
\DeclareOldFontCommand{\rm}{\normalfont\rmfamily}{\mathrm}

\usepackage{graphicx}      
\usepackage{subcaption}
\usepackage{enumitem}
\usepackage{mathdots}

\usepackage{mathtools,xparse}
\usepackage{epstopdf}
\def\interior{\operatorname{int}}
\def\bs{\boldsymbol}

\def\mcal{\mathcal}

\def\mcd{\mathcal{D}}
\def\mcn{\mathcal{N}}
\def\mcl{\mathcal{L}}

\def\I{\text{I}}
\def\S{\text{S}}
\def\DFE{\text{DFE}}

\newcommand{\seb}[1]{%
{\leavevmode\color{black}#1}%
}

\DeclareMathOperator{\diag}{\textrm{diag}}
\DeclareMathOperator{\spec}{spec}
\def\qed{\hfill $\Box$}

\begin{document}
\author{Benjamin Catalano, Keith Paarporn, and Sebin Gracy
\thanks{Benjamin Catalano and Sebin Gracy  are with the Department of Electrical  Engineering and Computer Science, South Dakota School of Mines, Rapid City, SD,USA (\texttt{benjamin.catalano@mines.sdsmt.edu}, \break \texttt{sebin.gracy@sdsmt.edu}). Keith Paarporn is with the Department of Computer Science at the University of Colorado-Colorado Springs, Co, USA \texttt{kpaarpor@uccs.edu}  
}.}
\title{\LARGE \bf Game-theoretic Social Distancing in  Competitive Bi-Virus SIS Epidemics}

\maketitle
\begin{abstract}
    Numerous elements drive the spread of infectious diseases in complex real-world networks. Of particular interest is social behaviors that evolve in tandem with the spread of disease. Moreover, recent studies highlight the importance of understanding how multiple strains spread simultaneously through a population (e.g. Delta and Omicron variants of SARS-CoV-2). In this paper, we propose a bi-virus SIS epidemic model coupled with a game-theoretic social distancing behavior model. The behaviors are governed by replicator equations from evolutionary game theory. The prevalence of each strain impacts the choice of an individual to social distance, and, in turn, their behavior affects the spread of each virus in the SIS model. Our analysis identifies equilibria of the system and their local stability properties, which reveal several isolated fixed points with varying levels of social distancing. We find that endemic co-existence is possible only when the reproduction numbers of both strains are equal. Assuming the reproduction number for each virus is the same, 
    we identify suitable parameter regimes that give rise to lines of coexistence equilibria. Moreover, we also identify conditions for local exponential stability of said lines of equilibria.
    We illustrate our findings with several numerical simulations.
\end{abstract}

\section{Introduction}

The dynamics of infectious disease spread has been studied for centuries, and has perpetually been a highly active research area. 
The recent COVID-19 pandemic illuminated a broad unpreparedness for a severe outbreak of a novel infectious disease by 
exposing knowledge gaps when it comes to the prediction and mitigation of outbreaks. 
Contributing to the unpreparedness was an overall inability to anticipate the public's social reactions to a quickly spreading disease, as well as the emergence of new, more severe strains that simultaneously spread in  populations.

Understanding these dynamics requires building and analyzing new classes of models that feature a co-evolution between decision-making (e.g. individual social distancing) and the spread of infectious diseases through physical contacts \cite{funk2015nine,heesterbeek2015modeling}. Recent research efforts have incorporated game-theoretic frameworks in order to model the social behaviors of individuals during epidemics \cite{ye2021game,paarporn2023sis,satapathi2023coupled,saad2023dynamics,hota2023learning}. In a game-theoretic formulation, the perceived costs and benefits from taking or not taking social distancing actions are the basis of how individuals make such decisions. Importantly, these costs and benefits are linked to how widespread the disease currently is. In turn, individual decision-making has an impact on the disease spread, thus forming a feedback between the two processes. In many of these studies, dynamic social behavior is modeled by incorporating replicator equations, which are standard evolutionary game-theoretic tools \cite{sandholm2010population,weitz2016oscillating,quijano2017role,satapathi2024game}.

The primary thread of literature 
\seb{that leverages notions from game theory for studying problems in epidemiology}
features a single virus strain spreading through the population.  However, it is often the case that \emph{multiple} strains spread simultaneously. Indeed, there is a large body of work that studies the dynamics of bi-virus (i.e., two virus) epidemics, though in the absence of any social behaviors \cite{ye2022convergence,gracy2025modeling,janson2024competitive,sahneh2014competitive,ye2024competitive}. One exception is \cite{liu2017continuous}, which studied bi-virus epidemics over networks with a mechanistic model of human awareness, which is closest in spirit to our work. Central goals in these works are to understand under what conditions both strains die out, when they may co-exist in a stable endemic state, or in the case of competitive viruses, when one of them dies out and the other remains endemic.

In this paper, we examine a novel competitive bi-virus epidemic model in which individual social-distancing is driven by game-theoretic behavior. Specifically, we consider an individual's infection status to be either susceptible or infected with one of the two strains. An individual chooses to either social distance or not, where social distancing reduces contact rates with other individuals. 
We use replicator equations to model how these decisions change over time. In our game-theoretic formulation, individuals base their decisions on costs associated with the perceived risks from being exposed to either one of the strains, as well as the economic costs from social distancing (e.g. staying home). We describe this model 
with a system of five coupled ordinary differential equations.

The primary contributions of this paper are \seb{as follows}:  \seb{First, we propose} a novel 
bi-virus dynamical model with game-theoretic social distancing behavior;  \seb{second, we provide} a comprehensive identification of the system's fixed points; and  \seb{finally, we identify conditions for (in)stability of the various fixed points.}
We find that there can exist numerous isolated fixed points with varying levels of social distancing, and which ones are stable depend on the parameters lying in certain ranges. None of the isolated fixed points reflect a co-existence of both strains. 
The only outcomes that exhibit co-existence arise when the reproduction numbers \seb{(i.e., the number of infections caused by an infected individual in a completely susceptible population)}  of both strains are identical -- in this case, there exist line segments of co-existence fixed points. Here, which point the system converges \seb{to} depends on the initial conditions.


\vspace{-2mm}
\subsection*{Paper Outline}
The paper unfolds in the following fashion: We conclude the present section by listing the notations  that will be used in the sequel. We introduce our model in Section~\ref{sec:model}. Sections~\ref{sec:analysis}, \ref{sec:stability:DFE:FPs}, and \ref{sec:line} deal with identifying the various fixed points of our system, identifying conditions for local stability/instability of DFE and the unilateral FPs, and securing conditions for local asymptotic stability of different lines of coexistence equilibria, respectively. We highlight our theoretical findings via numerical examples in Section~\ref{sec:sims}. Finally, we summarize our findings 
in Section~\ref{sec:conclusions}.

\vspace{-2mm}
\subsection*{Notations}
Let $\mathbb{R}$ and $\mathbb{R}_+$ denote the set of real numbers and the set of nonnegative real numbers, respectively.
Given a positive integer $n$, $[n]$ denotes the set $\{1,2,...,n\}$.
We denote logical conjunction and disjunction by $\land$ and $\lor$, respectively.
In the interest of conciseness, we refrain from specifying the dimensions of vectors/matrices unless these are not clear from the context.
Given a vector $x$, the square matrix with the elements of $x$ along the diagonal is denoted by $\diag{(x)}$. Given a matrix $A$, $A_{ij}$ denotes the element in the $i^{th}$ row and the $j^{th}$ column of $A$. We denote the $k$th row of the matrix $A$ by $A_{k,:}$.
For a set $S$, we denote the boundary of $S$ by $\partial S$ and the interior of $S$ by $\interior S$.
We denote the spectrum of matrix $A$ by $\spec(A)$. Suppose that $M \in \mathbb{R}^{n \times n}$, then $\spec(M)$ and $\rho(M)$ denote the spectrum of $M$, 
and the spectral radius of $M$, respectively. The spectral abscissa of $M$ is denoted by $s(M)$, i.e., $s(M) = \max\{\rm{Re}(\lambda) : \lambda \in \spec(M)\}$. 

\section{Model}\label{sec:model}

We consider a single well-mixed population of unit mass. There are two distinct competitive viruses spreading through the population, labeled virus 1 and 2. Each individual is either infected with strain 1, infected with strain 2, or susceptible. At any time $t \geq 0$, let us denote $s(t) \in [0,1]$ as the mass of susceptible individuals, $y_1(t) \in [0,1]$ 
as the mass of infected individuals with virus 1, and $y_2(t) \in [0,1]$ as the mass of infected individuals with virus 2. We use $\bs{y}(t) := [y_1(t), y_2(t)]^\top$ to denote the vector of infected masses. Together, these quantities must obey the conservation of mass, $s(t) + y_1(t) + y_2(t) = 1$. The virus strains have contact spreading rates of $\beta_1,\beta_2$, and infected individuals independently heal from the strains at rates $\delta_1,\delta_2$. 

\begin{assumption}\label{assume:betadelta}
    We assume the spreading rates satisfy $\beta_1,\beta_2 > 0$, and the healing rates satisfy $\delta_1,\delta_2 > 0$.
\end{assumption}

Imposing Assumption \ref{assume:betadelta} is to focus attention on the bi-virus dynamics -- setting one or both of the $\beta$ rates to zero reduces the analysis to a single-virus or no virus at all.

Individuals choose to either follow social distancing or not (action $\mcd$ or $\mcn$, resp.). We let $x_{\S\mcd}(t) \in [0,s(t)]$ and $x_{\S\mcn}(t) = s(t) - x_{\S\mcd}(t)$ denote the mass of susceptible individuals that follow and do not follow social distancing, respectively. Likewise, we denote $x_{i\mcd}(t) \in [0,y_i(t)]$ and $x_{i\mcn}(t) = y_i(t) - x_{i\mcd}(t)$ as the mass of individuals infected with virus $i=1,2$ that follow and do not follow social distancing, respectively.
The virus infections evolve according to the following dynamics:
\begin{equation}
    \begin{aligned}
        \dot{y}_1 &= \beta_1(qx_{\S\mcd} + x_{\S\mcn})(qx_{1\mcd} + x_{1\mcn}) - \delta_1 y_1 \\
        \dot{y}_2 &= \beta_2(qx_{\S\mcd} + x_{\S\mcn})(qx_{2\mcd} + x_{2\mcn}) - \delta_2 y_2 \\
    \end{aligned}
\end{equation}

Here, $q$ is the interaction reduction factor due to social distancing behavior.
\begin{assumption}\label{assume:q}
    We assume $q \in (0,1)$.
\end{assumption}
Low values of $q$ means more isolation, high values of $q$ means less isolation. 

\vspace{-2mm}
\subsection{Payoff functions}

The incentives to choose whether to social distance is modeled using payoff functions. For susceptible individuals, the perceived payoff for choosing to social distance is given by
\begin{equation}
    \pi_{\S,\mcd}(\bs{y}) := -c_\mcd + r_1 y_1(t) + r_2 y_2(t)
\end{equation}
where $c_\mcd > 0$ is the economic and social cost of taking social distancing measures. Individuals do not know the true probability of getting infected, but are typically informed about the total amount of people currently infected. Thus, we have defined the parameters $r_i > 0$ as perceived risk factors to being exposed to virus $i = 1,2$. Consequently, the perceived payoff for social distancing is increasing in the mass of infected individuals of either virus type. 

\begin{assumption}\label{assume:r}
    We will assume that $0 < r_1 < r_2$.
\end{assumption}

This assumption asserts that virus 1 is  perceived among the population to not be as severe as virus 2. For infected individuals of virus $i=1,2$, the perceived payoffs for choosing social distancing or not are given by

The perceived payoff for choosing to not social distance is
\begin{equation}
    \pi_{\S\mcn}(\bs{y}) := - (r_1 y_1(t) + r_2 y_2(t)).
\end{equation}
By not social distancing, the individual does not pay the cost $c_\mcd$, but pays a cost based on the perceived risks.

Infected individuals that social distance pay the cost $c_\mcd$ from before. Those that do not social distance pay a perceived cost $c_i > c_\mcd$ for putting other individuals at risk to virus $i$. We then define
\begin{equation}
        \pi_{i\mcd}(\bs{y}) := -c_\mcd \quad \text{and} \quad
        \pi_{i\mcn}(\bs{y}) := -c_i
\end{equation}

%
\normalsize

\begin{assumption}\label{assume:c}
    We will assume $c_i > c_\mcd$ for $i=1,2$.
\end{assumption}
This assumption asserts that the cost of socializing while infected is higher than the cost associated with social distancing. This scenario is plausible when either local authorities implement strict lockdown policies, or when a community's social norms discourage social activity when sick.

\vspace{-2mm}
\subsection{Coupled evolutionary dynamics}

We will use replicator equations to describe the evolution of social distancing behaviors. 
Define $z_\S(t) := x_{\S\mcd}(t)/s(t)$ as the \emph{fraction} of individuals among the susceptible population that social distance. Likewise, define $z_1(t) := x_{1\mcd}(t)/y_1(t)$ and $z_2(t) := x_{2\mcd}(t)/y_2(t)$. The replicator equation
\begin{equation}
    \dot{z}_\S = z_\S(1-z_\S)(\pi_{\S\mcd}(\bs{y}) - \pi_{\S\mcn}(\bs{y}))
\end{equation}
describes the evolution of social distancing behaviors among the susceptible population. Likewise, we also have replicator equations for the two infected subpopulations,
\begin{equation}
    \begin{aligned}
        \dot{z}_1 &= z_1(1-z_1)(\pi_{1\mcd}(\bs{y}) - \pi_{1\mcn}(\bs{y})) \\
        \dot{z}_2 &= z_2(1-z_2)(\pi_{2\mcd}(\bs{y}) - \pi_{2\mcn}(\bs{y}))
    \end{aligned}
\end{equation}
These dynamics give rise to a 5-dimensional system with state vector $\bs{p} = [y_1, y_2, z_\S, z_1, z_2]^\top$. The full set of coupled equations is:
\begin{equation}\label{eq:main}
    \begin{aligned}
        \dot{y}_1 &= y_1 (\beta_1 s (1 - z_\S(1-q))(1 - z_1(1-q)) - \delta_1) \\
        \dot{y}_2 &= y_2 (\beta_2 s (1 - z_\S(1-q))(1 - z_2(1-q)) - \delta_2) \\
        \dot{z}_\S &= z_\S(1-z_\S)(2(r_1y_1 + r_2y_2) - c_\mcd) \\
        \dot{z}_1 &= z_1(1-z_1)(c_1 - c_\mcd) \\
        \dot{z}_2 &= z_2(1-z_2)(c_2 - c_\mcd) 
    \end{aligned}
\end{equation}
Here, we have used the fact that each of the states $x_{\S\mcd}$, $x_{\I_1\mcd}$, and $x_{\I_2\mcd}$ determine the other states $x_{\S\mcn} = s(t) - x_{\S\mcd}$, $x_{\I_1\mcn} = y_1 - x_{\I_1\mcd}$, and $x_{\I_2\mcn} = y_2 - x_{\I_2\mcd}$.

We define the sets $\Delta$ and $\Gamma$ as follows:
\begin{align}
\Delta &:= \left\{(y_1,y_2) \;\big|\; \sum y_i \le 1 \land 0 \le y_i, i\in[2]\right\}\label{eq:Delta} \\
\Gamma &:= \Delta\times[0,1]^3 \label{eq:gamma}
\end{align}

\begin{lemma}\label{lem:pos_inv}
  \seb{Consider system~\eqref{eq:main} under Assumption~\ref{assume:betadelta}. The set $\Gamma$, where $\Gamma$ is as defined in~\eqref{eq:gamma}, is positively invariant.} 
\end{lemma}
\textit{Proof:} Observe that $\Gamma$ is a closed set, so $\partial\Gamma\subset\Gamma$ and it is impossible for a continuous trajectory to leave $\Gamma$ without passing through $\partial\Gamma$.
Therefore, we examine the behavior of system~\eqref{eq:main} at $\partial\Gamma$.
Consider $y_1 = 0$ (resp. $y_2 = 0$), then $\dot y_1 = 0$ (resp. $\dot y_2 = 0$). If $y_1 + y_2 = 1$, then $\dot y_i = -\delta_i y_i$, for $i\in[2]$.  
\seb{Due to Assumption~\ref{assume:betadelta}}, $\delta_i > 0, i\in[2]$. Therefore, it follows that at the boundary, $\partial\Delta$, $y_1$ (resp. $y_2$) is either at the lower bound, $0$, and constant or at the upper bound, $y_1 + y_2 = 1$, and decreasing.\\
Consider $z\in\{z_\S, z_1, z_2\}$. In each case, $\dot z$ 
depends on
$z(1-z)$. Since $z\in[0,1]$, the boundary values are given by $\partial[0,1]=\{0,1\}$, thus, evaluating the right hand side of the last three lines of~\eqref{eq:main} at each of the values in $\partial[0,1]$, we get $\dot z = 0$. Hence, $z_\S$, $z_1$ and $z_2$ are always constant 
at their respective boundaries.\\
Therefore, \seb{defining $x(t^\prime)=[\begin{smallmatrix}
    y_1(t^\prime) & y_2(t^\prime) &z_s(t^\prime) & z_1(t^\prime) &z_2(t^\prime)
\end{smallmatrix}]^\top$}, it must be that if $x(t_0)\in\Gamma$ then $x(t)\in\Gamma, \forall t > t_0$; that is, $\Gamma$ is positively invariant.~$\square$
\\
\seb{We need the following assumption to ensure that our model is well-defined.
\begin{assumption}\label{assume:init}
 $x(0):=[\begin{smallmatrix}
    y_1(0) & y_2(0) &z_s(0) & z_1(0) &z_2(0)
\end{smallmatrix}]^\top$. We have that $x(0)\in \Gamma$.   
\end{assumption}
In view of Assumption~\ref{assume:init}, Lemma~\ref{lem:pos_inv} guarantees that states  always take values in the $[0,1]$ interval. Note that if the states were to take values outside the $[0,1]$ interval, then those values will not correspond to physical reality.
}


\section{Analysis: Identification of fixed points}
\label{sec:analysis}
In this section, we identify all the fixed points (FPs) of 
system \eqref{eq:main}. From  Assumption \ref{assume:c} ($c_i > c_\mcd$), we have $\dot{z}_i > 0$ at any state in $\interior\Gamma$. In other words, infected individuals never have an incentive to not practice social distancing. As a result, any fixed point with $z_i = 0$ for any $i=1,2$ cannot be locally asymptotically stable. Therefore, we will rule these out and restrict attention to fixed points of the form $(y_1,y_2,z_S,1,1)$, with $y_1,y_2,z_S \in [0,1]$. 

We classify fixed points into the following three categories.
\begin{itemize}
    \item A \emph{disease-free equilibrium} (DFE) is any fixed point for which $y_1 = y_2 = 0$.
    \item A \emph{unilateral equilibrium} is any fixed point for which either $y_1 = 0$ and $y_2 > 0$, or $y_1 > 0$ and $y_2 = 0$.
    \item A \emph{coexistence equilibrium} is any fixed point for which $y_1,y_2 > 0$.
\end{itemize}

In order to characterize the full set of fixed points, we make use of the following functions.
\begin{align}
    \begin{aligned}
        h_1(y_1,y_2,z_\S,z_1) &:= - \delta_1+ \beta_1(1-y_1-y_2) \times \nonumber \\ &(qz_\S + (1-z_\S))(qz_1 + (1-z_1))   \\
        h_2(y_1,y_2,z_\S,z_2) &:= - \delta_2+ \beta_2(1-y_1-y_2) \times \nonumber \\  &(qz_\S + (1-z_\S))(qz_2 + (1-z_2))   \\
        h_\S(y_1,y_2) &:= 2(r_1y_1+r_2y_2)-c_\mcd
    \end{aligned}
\end{align}
\begin{lemma}[\textbf{Disease-Free Equilibria (DFE)}]\label{lem:DFE}
    There always exists two disease-free equilibria, $\bs{p}_{\DFE0} := [0,0,0,1,1]$ and $\bs{p}_{\DFE1} := [0,0,1,1,1]$.
\end{lemma}
\textit{Proof:}
    Setting $y_1 = y_2 = 0$, it only remains to solve $\dot{z}_\S = z_\S(1-z_\S)h_\S(0,0) = 0$. It cannot be the case that $h_\S(0,0) = 0$ since $h_\S(0,0) = -c_\mcd < 0$. This yields the two DFEs.~$\square$
%

In the next Lemma, we identify all unilateral equilibria in the system \eqref{eq:main}, as well as conditions on the parameters for which each one lies in the state space $\Gamma$.

\begin{lemma}[\textbf{Unilateral Equilibria}]\label{lem:unilateral}
    The following is a characterization of all the unilateral equilibria in system \eqref{eq:main} and their existence conditions. For $i=1,2$,

    \noindent 1) The equilibrium $\bs{p}_{i0}$, defined by $y_i = 1 - \frac{\delta_i}{q\beta_i}$, $y_{3-i} = 0$, $z_S = 0$, and $z_1=z_2=1$, exists if and only if $\frac{\delta_i}{q\beta_i} < 1$.

    \noindent 2) The equilibrium $\bs{p}_{i1}$, defined by $y_i = 1 - \frac{\delta_i}{q^2\beta_i}$, $y_{3-i} = 0$, $z_S = 0$, and $z_1=z_2=1$, exists if and only if $\frac{\delta_i}{q^2\beta_i} < 1$.

    \noindent 3) The equilibrium $\bs{p}_{i\S}$, defined by $y_i = \frac{c_\mcd}{2r_i}$, $y_{3-i} = 0$, $z_S = \frac{1}{1-q} - \frac{\delta_i}{\beta_i(1-\frac{c_\mcd}{2r_i})q(1-q)}$, and $z_1=z_2=1$, exists if and only if $\frac{c_\mcd}{2r_i} \leq 1$ and
    \begin{equation}\label{eq:p1S_condition}
        q (1-\frac{c_\mcd}{2r_i}) < \frac{\delta_i}{q\beta_i} < (1-\frac{c_\mcd}{2r_i}).
    \end{equation}
\end{lemma}
\textit{Proof:} We focus on proving the case $i=1$, as the case $i=2$ will follow completely analogous arguments. Thus, our task is to identify all fixed points with $y_1 > 0$ and $y_2 = 0$. We prove each part separately.
    
    \noindent(1) In this part, we suppose that $z_S = 0$, which sets $\dot{z}_S = 0$. We then need to solve $h_1(y_1,0,0,1) = \beta_1(1-y_1) q - \delta_1 = 0$, resulting in $y_1 = 1 - \frac{\delta_1}{q\beta_1}$. It holds that $y_1 > 0$ if and only if $\frac{\delta_i}{q\beta_i} < 1$. 

    \noindent(2) In this part, we suppose that $z_S = 1$, which gives $\dot{z}_S = 0$. We the need to solve $h_1(y_1,0,1,1) = \beta_1(1-y_1) q^2 - \delta_1 = 0$, resulting in $y_1 = 1 - \frac{\delta_1}{q^2\beta_1}$. It holds that $y_1 > 0$ if and only if $\frac{\delta_i}{q^2\beta_i} < 1$.

    \noindent(3) In this part, we suppose that $z_S \in (0,1)$. In order for $\dot{z}_S = 0$, we need that $h_S(y_1,0) = 0$, which gives $y_1 = \frac{c_\mcd}{2r_1}$. For $y_1 \in (0,1]$, it is required that $\frac{c_\mcd}{2r_1} \leq 1$. In order for $\dot{y}_1 = 0$, we need to solve $h(\frac{c_\mcd}{2r_1},0,z_S,1) = 0$, which yields $z_S = \frac{1}{1-q} - \frac{\delta_1}{\beta_1(1-\frac{c_\mcd}{2r_1})q(1-q)}$. For $z_S > 0$, it is required that $\frac{\delta_i}{q\beta_i} < (1-\frac{c_\mcd}{2r_i})$. For $z_S < 1$, it is required that $q (1-\frac{c_\mcd}{2r_i}) < \frac{\delta_i}{q\beta_i}$.~$\square$

The equilibrium $\bs{p}_{i0}$ indicates a unilateral endemic state in which virus $i$ survives, and nobody in the population is social distancing. The equilibrium $\bs{p}_{i1}$ indicates a unilateral endemic state in which virus $i$ survives and everybody in the population is social distancing. The equilibrium $\bs{p}_{iS}$ indicates a unilateral endemic state in which virus $i$ survives and a fraction of the population is social distancing.

In the next Lemma, we identify the set of all coexistience equilibria and conditions for when they exist in the state space $\Gamma$. 

\begin{lemma}[\textbf{Coexistence equilibria}]\label{lem:coexist}
    Coexistence equilibria in system \eqref{eq:main} can exist only if \seb{the reproduction number $R_0$ is such that} $R_0 := \frac{\beta_1}{\delta_1} = \frac{\beta_2}{\delta_2}$. We characterize all such equilibria below.

    \vspace{1mm}
    
    \noindent 1) A line of coexistence equilibria of the form
    \begin{equation}\label{eq:line0}
        \mathcal{L}_0 := \left\{ (y_1,y_2,0,1,1) : y_1 + y_2 = 1 - \frac{1}{q R_0} \right\},
    \end{equation}
    exists if and only if $qR_0 > 1$.

    \vspace{1mm}
    
    \noindent 2) A line of equilibria of the form
    \begin{equation}\label{eq:line1}
        \mathcal{L}_1 := \left\{ (y_1,y_2,1,1,1) : y_1 + y_2 = 1 - \frac{1}{q^2R_0} \right\}
    \end{equation}
    exists if and only if $q^2 R_0 > 1$.

    \vspace{1mm}
    
    \noindent 3) Denote $\mcal{L}_S$ as the set of points of the form $(y_1,y_2,z_S,1,1)$ that is parameterized by the value $y_1$, where $y_2 = \frac{c_\mcd}{2r_2} - \frac{r_1}{r_2}y_1$,  $z_S = \frac{1}{1-q} - \frac{1}{q(1-q)R_0(1-y_1-y_2)}$, and $y_1$ lies in the range
    \begin{equation}\label{eq:lineS}
        \underline{B} < y_1 < \bar{B},
    \end{equation}
    where
    \begin{equation}
        \underline{B}:=\max\left\{ 0, \frac{r_2}{r_1}\left( \frac{c_\mcd}{2r_2} - 1 \right), \ \frac{1 - \frac{c_\mcd}{2r_2} - \frac{1}{q^2R_0}}{1 - \frac{r_1}{r_2}} \right\}
    \end{equation}
    and
    \begin{equation}
        \bar{B} := \min\left\{ \frac{c_\mcd}{2r_1}, \ \frac{1 - \frac{c_\mcd}{2r_2}}{1 - \frac{r_1}{r_2}}, \ \frac{1 - \frac{c_\mcd}{2r_2} - \frac{1}{qR_0}}{1 - \frac{r_1}{r_2}}, 1 \right\}.
    \end{equation}
    Then $\mcal{L}_S$ is a line of equilibria contained in the state space $\Gamma$ if and only if $\underline{B} < \bar{B}$.
\end{lemma}
\textit{Proof:}
    We prove each case separately.

    \noindent(1) Suppose that $z_S = 0$. For $\dot{y}_1 = 0$, we solve $h_1(y_1,y_2,0,1) = \beta_1(1-y_1-y_2) \cdot q - \delta_1 = 0$, resulting in $y_1 + y_2 = 1 - \frac{\delta_1}{q\beta_1}$. For $\dot{y}_2 = 0$, we solve $h_2(y_1,y_2,0,1) = 0$, resulting in $y_1 + y_2 = 1 - \frac{\delta_2}{q\beta_2}$. In order for a fixed point with $y_1,y_2 > 0$ satisfying these equalities to exist, it is required that $\frac{\delta_1}{\beta_1} = \frac{\delta_2}{\beta_2}$. Then, the line of equilibria $\mcal{L}_0$ exists if and only if $0<1-\frac{1}{qR_0} < 1$. Since $R_0, q > 0$ by assumptions \ref{assume:betadelta} and \ref{assume:q}, this is equivalent to $qR_0 > 1$.

    \noindent(2) This case is analogous to case (1), and so we omit these details for brevity.

    \noindent(3) Here, suppose that $z_\S \in (0,1)$. For $\dot{z}_S = 0$, it holds that $h_S(y_1,y_2) = 0$, or that $y_2 = \frac{c_\mcd}{2r_2} - \frac{r_1}{r_2}y_1$. In order for $y_2 \in (0,1)$, it must hold that
    \begin{equation}\label{eq:coex_range1}
        \frac{r_2}{r_1}\left( \frac{c_\mcd}{2r_2} - 1 \right) < y_1 <    \frac{c_\mcd}{2r_1}.
    \end{equation}
    
    Let us denote $\ell(y_1) := 1 - y_1 - y_2 = (1-\frac{c_\mcd}{2r_2}) - y_1(1-\frac{r_1}{r_2})$. Note it must hold that $\ell(y_1) \geq 0$, or 
    \begin{equation}\label{eq:coex_range2}
        y_1 \leq \frac{1 - \frac{c_\mcd}{2r_2}}{1 - \frac{r_1}{r_2}}.
    \end{equation}

    For $\dot{y}_1 = 0$, we solve $h_1(y_1,y_2,z_S,1) = 0$, which yields $z_\S = \frac{1}{1-q} - \frac{\delta_1}{q(1-q)\beta_1 \ell(y_1)}$. Similarly, for $\dot{y}_2 = 0$, we solve $h_2(y_1,y_2,z_S,1) = 0$, which yields $z_\S = \frac{1}{1-q} - \frac{\delta_2}{q(1-q)\beta_2 \ell(y_1)}$. For these two equations to be satisfied, it is required that $\frac{\delta_1}{\beta_1} = \frac{\delta_2}{\beta_2}$.

    Now, the condition that $z_S \in (0,1)$ is equivalent to
    \begin{equation}\label{eq:coex_range3}
        \frac{1 - \frac{c_\mcd}{2r_2} - \frac{1}{q^2R_0}}{1 - \frac{r_1}{r_2}} < y_1 < \frac{1 - \frac{c_\mcd}{2r_2} - \frac{1}{qR_0}}{1 - \frac{r_1}{r_2}}.
    \end{equation}

    Putting together all conditions on $y_1$ \eqref{eq:coex_range1}, \eqref{eq:coex_range2}, and \eqref{eq:coex_range3}, in addition to the restriction $y_1 \in (0,1)$, we conclude that a line of equilibria defined by $\mcal{L}_S$ lies in the state space $\Gamma$ if and only if $\underline{B} < \bar{B}$. ~$\square$\\   
%
Lemma \ref{lem:coexist} asserts that no coexistence equilibrium can be an isolated fixed point. They must always exist either no coexistence equilibria, or an infinite number of them situated on line(s) of coexistence fixed points. We note that the conditions for the existence of the three lines are not mutually exclusive. Also, we remark that coexistence fixed points can only exist when the reproduction numbers of both viruses are identical, i.e. $R_0 = \beta_1/\delta_1 = \beta_2/\delta_2$.
Interestingly, this necessary condition does not depend on the risk perception parameters $r_1,r_2$.


\section{Stability analysis of DFE and unilateral FPs}\label{sec:stability:DFE:FPs}
In this section, we identify parameter-based conditions first for  stability/instability of $\bs{p}_{\DFE0}$ and $\bs{p}_{\DFE1}$, and subsequently for unilateral FPs,  $\bs{p}_{10}$, $\bs{p}_{11}$, and $\bs{p}_{1S}$.   
\vspace{-3mm}
\subsection{Stability analysis of DFE}
In this subsection, we secure a condition for local exponential stability of  $\bs{p}_{\DFE0}$ and then show that the FP $\bs{p}_{\DFE1}$ is \emph{never} stable. We 
have the following result.
\begin{proposition}\label{prop:localstability:DFE0}
   \seb{Consider system~\eqref{eq:main} under Assumptions~\ref{assume:betadelta}, \ref{assume:q}\ and \ref{assume:c}}. The fixed point $\bs{p}_{\DFE0} = [0,0,0,1,1]$ is locally exponentially stable if $\delta_k>q\beta_k$ for $k=1,2$. If, for some $k \in [2]$, $\delta_k < q\beta_k$, then the fixed point $\bs{p}_{\DFE0} = [0,0,0,1,1]$  is unstable.
\end{proposition}
\noindent \textit{Proof:} Observe that the Jacobian of system~\eqref{eq:main}, evaluated at $\bs{p}_{\DFE0}= [0,0,0,1,1]$ (referred to as $J(0,0,0,1,1)$) is a diagonal matrix; the elements along the diagonal are $\beta_1 q - \delta_1$, $\beta_2 q - \delta_2$, $-c_\mcd$, $-(c_1-c_\mcd)$, and $-(c_2-c_\mcd)$, which are also the eigenvalues of  $J(0,0,0,1,1)$. 
By assumption, $\delta_k>q\beta_k$ for $k=1,2$. \seb{This, since by Assumption~\ref{assume:betadelta}, $\beta_k>0, \delta_k>0$, and since by Assumption~\ref{assume:q}, $q \in (0,1]$}, implies that $q\beta_k-\delta_k<0$ for $k=1,2$. \seb{By Assumption~\ref{assume:c}, we know that} a) $c_\mcd>0$, b) $c_1>c_\mcd$, and c) $c_2>c_\mcd$. Therefore, it is straightforward to see that $-c_\mcd<0$, $-(c_1-c_\mcd)<0$, and $-(c_2-c_\mcd)<0$; in view of the discussion above, this means that all of the eigenvalues of $J(0,0,0,1,1)$ are (real and) negative. Hence, $s(J(0,0,0,1,1))<0$. Local exponential stability of the fixed point  $\bs{p}_{\DFE0} = [0,0,0,1,1]$, then, follows from \cite[Theorem 4.15 and Corollary~4.3]{khalil2002nonlinear}.\\
Suppose that, for some $k \in [2]$, $\delta_k < q\beta_k$. Then it is clear that at least one (possibly two) eigenvalue of $J(0,0,0,1,1)$ is positive, which means that $s(J(0,0,0,1,1))>0$. Consequently, instability of $\bs{p}_{\DFE0}= [0,0,0,1,1]$ follows from \cite[Theorem~4.7, statement ~ii)]{khalil2002nonlinear}.~\qed

We have the following remark.
\begin{remark}\label{rem:p_DFE0}[Epidemiological interpretation] From an epidemiological  viewpoint, Proposition~\ref{prop:localstability:DFE0} says that as long as the healing rate is larger than the scaled (by the value of the interaction reduction factor) infection rate, then, assuming that the initial infection levels with respect to the two viruses, are close enough to  $\bs{p}_{\DFE0}$, the two viruses gets eradicated, and, quite naturally, none of the individuals in the population node practises social distancing.     
\end{remark}


We next turn our attention to the stability (or lack thereof) of the fixed point $\bs{p}_{\DFE1} = [0,0,1,1,1]$. We have the following result.
\begin{proposition}\label{prop:dfe1}
    \seb{Consider system~\eqref{eq:main} under Assumption~\ref{assume:betadelta} and \ref{assume:c}}. The fixed point $\bs{p}_{\DFE1} := [0,0,1,1,1]$ is always unstable.
\end{proposition}
\noindent \textit{Proof:} It is straightforward to show that the Jacobian of system~\eqref{eq:main}, evaluated at $\bs{p}_{\DFE1} = [0,0,1,1,1]$ (referred to as $J(0,0,1,1,1)$) is a diagonal matrix; the elements along the diagonal are $\beta_1 q^2-\delta_1, \beta_2 q^2-\delta_2, c_\mcd, -(c_1-c_\mcd)$ and $-(c_2-c_\mcd)$, which are also the eigenvalues of $J(0,0,1,1,1)$.
\seb{By Assumption~\ref{assume:c}, we know} that $c_\mcd >0$. Therefore, regardless of values that, for $k=1,2$, $\beta_k, \delta_k$   and $q$ take, the matrix $J(0,0,1,1,1)$ is never Hurwitz, since $s(J(0,0,1,1,1))>0$. Instability of $\bs{p}_{\DFE1} := [0,0,1,1,1]$, then, follows from from \cite[Theorem~4.7, statement ~ii)]{khalil2002nonlinear}.~\qed

We have the following remark.
\begin{remark}\label{rem:p_DFE1} Proposition~\ref{prop:dfe1} says that if both viruses are extinct (i.e., $y_1=y_2=0$), then, irrespective of the healing and infection rates,  it does not make sense for individuals to practise social distancing. Hence, the equilibrium point   $\bs{p}_{\DFE1}$ acts as a repeller; it drives the solution trajectories of system~\eqref{eq:main} away from it.    
\end{remark}

\vspace{-2mm}

\subsection{Stability analysis of unilateral FPs}
First, we investigate the stability of the unilateral FP, $\bs{p}_{10} = (1 - \frac{\delta_1}{q\beta_1},0,0,1,1)$. 
We have the following result.

\begin{proposition}\label{prop:fp1:local stability}
   \seb{Consider system~\eqref{eq:main} under Assumptions~\ref{assume:betadelta} and~\ref{assume:c}. Suppose that $1 < q\frac{\beta_1}{\delta_1}$}. The equilibrium point $\bs{p}_{10} = (1 - \frac{\delta_1}{q\beta_1},0,0,1,1)$ is locally exponentially stable if each of the following condition is satisfied:
   \begin{enumerate}[label=\roman*)]
       \item\label{p01} $1 < q\frac{\beta_1}{\delta_1}$;
       \item\label{p02} $1 > q\frac{\beta_2}{\delta_2}$;
       \item\label{p03} $\frac{c_\mcd}{2r_1} > (1-\frac{\delta_1}{q\beta_1})$
   \end{enumerate}
   If $\delta_2<\beta_2 q$ and/or if $c_\mcd<2r_1(1-\frac{\delta_1}{q\beta_1})$, then $\bs{p}_{10}$  is unstable.
\end{proposition}
\noindent \textit{Proof:} 
\seb{By assumption, $\frac{\delta_1}{q\beta_1}<1$, which from Lemma~\ref{lem:unilateral} item~i) ensures the
existence of fixed point $\bs{p}_{10}$.}
The Jacobian evaluated at $\bs{p}_{10}$ 
reads as in~\eqref{eq:jacobian:p10}.
\begin{figure*}
\begin{equation}\label{eq:jacobian:p10}
J(\bs{p}_{10}) = \scriptsize
\begin{bmatrix}
    \delta_1 - \beta_1 q &
    \delta_1 - \beta_1 q &
    -\delta_1 (1 - \frac{\delta_1}{q\beta_1}) (1 - q) &
    -\delta_1 (1 - \frac{\delta_1}{q\beta_1}) \frac{(1 - q)}{q} &
    0 \\
    0 & \beta_2 \frac{\delta_1}{\beta_1} - \delta_2 & 0 & 0 & 0 \\
    0 & 0 & 2 r_1 (1 - \frac{\delta_1}{q\beta_1}) - c_\mathcal{D} & 0 & 0 \\
    0 & 0 & 0 & -(c_1 - c_\mathcal{D}) & 0 \\
    0 & 0 & 0 & 0 & -(c_2 - c_\mathcal{D}) \\
\end{bmatrix}
\end{equation}
\end{figure*}
Note that $ J(\bs{p}_{10})$ is upper triangular; its eigenvalues are the entries along its diagonal. Since, by assumption, $\delta_1<q\beta_1$, it is clear that $ J(\bs{p}_{10})_{11}<0$.
Observe that $J(\bs{p}_{10})_{22} < 0 \iff \frac{\beta_1}{\delta_1} > \frac{\beta_2}{\delta_2}$. Since, by assumption $q\frac{\beta_2}{\delta_2} < 1 < q\frac{\beta_1}{\delta_1}$, we have $\frac{\beta_2}{\delta_2} < \frac{1}{q} < \frac{\beta_1}{\delta_1}$. Therefore, $J(\bs{p}_{10})_{22} < 0$.
The assumption $c_\mcd>2r_1(1-\frac{\delta_1}{q\beta_1})$ ensures that  $J(\bs{p}_{10})_{33}<0$.  Using Assumption~\ref{assume:c}, the rest of the proof is similar to the proof of Proposition~\ref{prop:localstability:DFE0}.~\qed

Note that an analogous result establishing local exponential stability of the fixed point $\bs{p}_{20} = (0, 1 - \frac{\delta_2}{q\beta_2},0,1,1)$ can be obtained by means of a suitable adjustment of notations.


Next, we focus on the stability of $\bs{p}_{11} = (1 - \frac{\delta_1}{q\beta_1},0,1,1,1)$. We provide a sufficient condition for local exponential stability of $\bs{p}_{11}$, and also identify multiple necessary conditions for the same. Our result is as follows. \begin{proposition}\label{prop:fp2:local stability}
   \seb{Consider system~\eqref{eq:main} under Assumptions~\ref{assume:betadelta}, \ref{assume:q}, and \ref{assume:c}. Suppose further that $\frac{\delta_1}{q^2\beta_1} < 1$}. The equilibrium point $\bs{p}_{11} = (1 - \frac{\delta_1}{q\beta_1},0,1,1,1)$ is locally exponentially stable if each of the following condition is satisfied:
   \begin{enumerate}[label=\roman*)]
       \item\label{p11} $1 < q^2\frac{\beta_1}{\delta_1}$;
       \item\label{p12} $1 > q^2\frac{\beta_2}{\delta_2}$;
       \item\label{p13} $\frac{c_\mcd}{2 r_1} < (1-\frac{\delta_1}{q^2\beta_1})$
   \end{enumerate}
   If $\delta_2 < \beta_2 q^2$ or $c_\mcd < 2r_1(1-\frac{\delta_1}{q\beta_1})$, then $\bs{p}_{11}$ is unstable.
\end{proposition}
\noindent \textit{Proof:} 
\seb{By assumption, $\frac{\delta_1}{q^2\beta_1} < 1$, which, from Lemma~\ref{lem:unilateral} item~ii), guarantees the existence of the FP $\bs{p}_{11}$.}
The Jacobian evaluated at $\bs{p}_{11}$ 
reads as in~\eqref{eq:jacobian:p11}. 
\begin{figure*}
\begin{equation}\label{eq:jacobian:p11}
J(\bs{p}_{11}) = \scriptsize
\begin{bmatrix}
\delta_1 - \beta_1 q^2 & \delta_1 - \beta_1 q^2 &
-\delta_1 (1 - \frac{\delta_1}{q^2\beta_1}) \frac{(1 - q)}{q} &
-\delta_1 (1 - \frac{\delta_1}{q^2\beta_1}) \frac{(1 - q)}{q} & 0 \\
0 & \beta_2 \frac{\delta_1}{\beta_1} - \delta_2 & 0 & 0 & 0 \\
0 & 0 & c_\mathcal{D} - 2 r_1 (1 - \frac{\delta_1}{q^2\beta_1}) & 0 & 0 \\
0 & 0 & 0 & -(c_1 - c_\mathcal{D}) & 0 \\
0 & 0 & 0 & 0 & -(c_2 - c_\mathcal{D}) \\
\end{bmatrix}
\end{equation}
\end{figure*}
Note that $J(\bs{p}_{11})$ is upper triangular; its eigenvalues are the entries along its diagonal.
Note that $J(\bs{p}_{11})_{11} = \delta_1 - \beta_1 q^2$. Hence, under assumption~\ref{p11} $J(\bs{p}_{11})_{11}$ is negative. Consider $J(\bs{p}_{11})_{22} = \beta_2\frac{\delta_1}{\beta_1} - \delta_2$. This is negative iff $\frac{\beta_1}{\delta_1} > \frac{\beta_2}{\delta_2}$. Under assumptions~\ref{p11} and ~\ref{p12}, we get the chain inequality $q^2\frac{\beta_2}{\delta_2} < 1 < q^2\frac{\beta_1}{\delta_1}$; therefore, $J(\bs{p}_{11})_{22} < 0$.
By assumption~\ref{p13}, it is clear that $J(\bs{p}_{11})_{33}<0$.
The rest of the proof is similar to the proof of Proposition~\ref{prop:localstability:DFE0} using Assumption~\ref{assume:c}.~\qed


Next, we identify sufficient conditions for the (in)stability of $\bs{p}_{1\S}$. We have the following result.
\begin{proposition}\label{prop:p1s}
     Consider system~\eqref{eq:main} under Assumptions~\ref{assume:betadelta} and \ref{assume:c}.
     If $c_\mcd < 2 r_i$, for each $i\in[2]$, then the fixed point $\bs{p}_{1\S}$ exists in $\Gamma$, and it is stable (resp. unstable) if $\beta_1/\delta_1 > \beta_2/\delta_2$ (resp. $\beta_1/\delta_1 < \beta_2/\delta_2$).
\end{proposition}
\textit{Proof:}
\seb{The assumption $c_\mcd < 2 r_i, i\in[2]$ is sufficient to satisfy the condition in  statement (iii) of Lemma~\ref{lem:unilateral}; thus, $\bs{p}_{1\S}$ is guaranteed to exist. The Jacobian evaluated at $\bs{p}_{1\S}$, post a suitable simplification, is as given in~\eqref{eq:jacobian:p1S}.}
\begin{figure*}
\begin{equation}\label{eq:jacobian:p1S}\scriptsize
J(\bs{p}_{1\S}) = \begin{bmatrix}
-\frac{\delta_1}{1-\frac{c_\mcd}{2r_1}} \cdot\frac{c_\mcd}{2r_1} &
-\frac{\delta_1}{1-\frac{c_\mcd}{2r_1}} \cdot\frac{c_\mcd}{2r_1} &
-\beta_1 \frac{c_\mcd}{2r_1} (1 - \frac{c_\mcd}{2r_1}) (1 - q) q &
-\delta_1\frac{c_\mcd}{2r_1} \frac{1 - q}{q} & 0 \\
0 & \frac{\beta_2}{\beta_1}\delta_1 - \delta_2 & 0 & 0 & 0 \\
z_S(1 - z_S) 2 r_1 & z_S(1 - z_S) 2 r_2 & 0 & 0 & 0\\
0 & 0 & 0 & -(c_1 - c_\mcd) & 0 \\
0 & 0 & 0 & 0 & -(c_2 - c_\mcd) \\
\end{bmatrix}
\end{equation}
\end{figure*}
We partition $J(\bs{p}_{1\S}) = [J_1(\bs{p}_{1\S}), J_2(\bs{p}_{1\S}); \bs{0}_{2\times3}, J_3(\bs{p}_{1\S})]$. Since this partitioning is upper triangular, the spectrum of the Jacobian is given by the spectra of $J_1(\bs{p}_{1\S})$ and $J_3(\bs{p}_{1\S})$.
Since $J_3(\bs{p}_{1\S})$ is diagonal, we immediately have the eigenvalues $\lambda_4 = -(c_1 - c_\mcd)$ and $\lambda_5 = -(c_2 - c_\mcd)$, which are both negative under Assumption~\ref{assume:c}.

Consider $\spec(J_1(\bs{p}_{1\S}))$. With $J_{ij} := [J_1(\bs{p}_{1\S})]_{ij}$, the characteristic equation of $J_1(\bs{p}_{1\S})$ is:
$$\det(\lambda\I - J_1(\bs{p}_{1\S}))
= (\lambda - J_{22})(\lambda(\lambda - J_{11}) - J_{13} J_{31}) = 0$$
Immediately, we have the eigenvalue $\lambda_2 = J_{22} = \frac{\beta_2}{\beta_1}\delta_1 - \delta_2$. Thus, \seb{by Assumption~\ref{assume:betadelta}}, $\lambda_2 < 0 \iff \frac{\beta_1}{\delta_1} > \frac{\beta_2}{\delta_2}$ and $\lambda_2 > 0 \iff \frac{\beta_1}{\delta_1} < \frac{\beta_2}{\delta_2}$.

Applying the quadratic formula to the remaining factor yields:
$$\{\lambda_1, \lambda_3\} = \frac{1}{2} J_{11} \pm \sqrt{J_{11}^2 + 4 J_{13} J_{31}}$$
Observe that under our Assumptions~\ref{assume:betadelta} and \ref{assume:c} with $c_\mcd < 2r_1$, we have $J_{11} < 0$, $J_{13} < 0$. From our earlier application of Lemma~\ref{lem:unilateral} (iii), we know $z_\S\in(0,1)$; thus, since $c_\mcd < 2r_1$ and so $r_1 > 0$, we have $J_{31} > 0$ (see Equation~\ref{eq:jacobian:p1S}).
Therefore, $4J_{13}J_{31} < 0$. Thus, if $|J_{11}^2| > |4J_{13} J_{31}|$ then $\left|\sqrt{J_{11}^2 + 4J_{13}J_{31}}\right| < |J_{11}|$, and the larger eigenvalue is bounded by: $\lambda < \frac{1}{2} \left[J_{11} + |J_{11}|\right] = 0$, i.e., both eigenvalues are negative. Alternatively, if $|J_{11}^2| < |4J_{13} J_{31}|$, then the quantity under the radical is negative and real. Therefore, its roots are purely imaginary, and, since $J_{11} < 0$, the result has strictly negative real part. Therefore, $\lambda_1$ and $\lambda_3$ are always negative. Consequently, $\bs{p}_{1\S}$ is stable if $\beta_1/\delta_1 > \beta_2/\delta_2$; it is 
unstable if $\beta_1/\delta_1 < \beta_2/\delta_2$.~$\square$

\section{Stability of line of coexistence equilibria} \label{sec:line}
In this section, we identify conditions 
for (in)stability of lines of coexistence equilibria, $\mathcal L_0$ and $\mathcal L_1$. \seb{We say that a line of equilibria is \emph{stable} when every point on that line is a stable equilibrium point; otherwise, we say it (i.e., the line of equilibria) is unstable.}
\vspace{-3ex}

\subsection{Stability of line $\mathcal L_0$}
Our first main result is the following theorem.
\begin{theorem}\label{thm:l0}
    \seb{Consider system~\eqref{eq:main} under Assumptions~\ref{assume:betadelta}, \ref{assume:q}, and 
    \ref{assume:c}. 
   Suppose further that $R_0 = \frac{\beta_1}{\delta_1} = \frac{\beta_2}{\delta_2}$ and $qR_0 > 1$.
    Then the line of equilibria $\mcl_0$ is locally exponentially stable if
    $1 - \frac{1}{qR_0} < \frac{c_\mathcal{D}}{2r_2}$ and $1 - \frac{1}{qR_0} < \frac{c_\mathcal{D}}{2r_1}$; it is unstable if $1 - \frac{1}{qR_0} > \frac{c_\mathcal{D}}{2r_2}$ or $1 - \frac{1}{qR_0} > \frac{c_\mathcal{D}}{2r_1}$.}
\end{theorem}
\textit{Proof:} Under the hypothesis of the theorem, the conditions in Lemma~\ref{lem:coexist}, statement (1) are satisfied,  and, consequently,  $\mcl_0$ is guaranteed to exist. Equation~\eqref{eq:line0} states that $1 - y_1 - y_2 = s = \frac{1}{qR_0}$, and so $\beta_isq = \delta_i, i\in[2]$. We use this identity, as well as substituting, $z_S = 0, z_1 = z_2 = 1$, to simplify the Jacobian, as given in~\eqref{eq:jacobian:line:l0}. %
\begin{figure*}\footnotesize
\begin{equation}\label{eq:jacobian:line:l0}     
J(\mcl_0) = \begin{bmatrix}
    -\beta_1 y_1 q & -\beta_1 y_1 q &
    -\beta_1 y_1 s (1 - q) q & -\beta_1 y_1 s (1 - q) & 0 \\
    -\beta_2 y_2 q & -\beta_2 y_2 q &
    -\beta_2 y_2 s (1 - q) q & 0 & -\beta_2 y_2 s (1 - q) \\
    0 & 0 & 2(r_1 y_1 + r_2 y_2) - c_\mathcal{D} & 0 & 0\\
    0 & 0 & 0 & -(c_1 -c_\mathcal{D}) & 0 \\
    0 & 0 & 0 & 0 & -(c_2 - c_\mathcal{D}) \\
\end{bmatrix}
\end{equation}
\end{figure*}
\normalsize
We partition $J(\mcl_0) = [J_1(\mcl_0), J_2(\mcl_0); \bs{0}_{3\times2}, J_3(\mcl_0)]$.
Observe that $J(\mcl_0)$ is a block upper triangular matrix, so its spectrum is given by the spectrum of $J_1(\mcl_0)$ and the spectrum of $J_3(\mcl_0)$.
Let $\lambda_i$ denote the $i^{th}$ eigenvalue of $J(\mcl_0)$.
Consider $J_1(\mcl_0)$, which has the structure $[a, a; b, b]$, where  $a = -\beta_1 y_1 q$ and $b = -\beta_1 y_2 q$. This yields the eigen pairs $\{(0, [1,-1]^T), (a+b,[a/b,1]^T\}$.
By Assumption~\ref{assume:betadelta}, $\beta_i,\delta_i > 0, i\in[2]$; with the assumption $qR_0 > 1$ we have $0 < \frac{1}{q R_0} < 1$. Then the line equation requires $y_1 + y_2 = 1 - \frac{1}{q R_0}\in(0,1)$. Therefore, since, by Assumption~\ref{assume:q}, $q \in (0,1]$, we have that $\lambda_2 = a + b = (-q(\beta_1 y_1 + \beta_2 y_2)) < 0$.
Next, we consider a perturbation along the direction of the eigenvector corresponding to $\lambda_1 = 0$: Let $\Delta \bs{y} := \epsilon\cdot[1, -1]^T$ for some arbitrarily small $|\epsilon|$. Observe that if $\bs{y}^*\in\mcl_0$, then $\bs{y}^* + \Delta \bs{y}\in\mathcal{L}_0$. This follows from substitution into the line equation:
$1 - y_1 - y_2 = \frac{1}{qR_0}
\iff 1 - (y_1 + \epsilon) - (y_2 - \epsilon) = \frac{1}{qR_0}$.
Therefore, the null-space of $J_1(\mcl_0)$ corresponds to the line of equilibria.

Consider $J_3(\mcl_0) = \diag(2(r_1 y_1 + r_2 y_2) - c_\mathcal{D}, c_1 - c_\mcd, c_2 - c_\mcd)$. Being a diagonal matrix, the eigenvalues of $J_3(\mcl_0)$ are its diagonal entries. 
Thus, $\lambda_3 = 2(r_1 y_1 + r_2 y_2) - c_\mathcal{D}$, $\lambda_4 = c_\mcd - c_1$ and $\lambda_5 = c_\mcd - c_2$. Since by Assumption~\ref{assume:c}, $c_\mcd < c_i, i\in[2]$, we have $\lambda_4 < 0$ and $\lambda_5 < 0$.
Consider $\lambda_3$. For it to be negative, we require that $2(r_1 y_1 + r_2 y_2) - c_\mathcal{D} < 0$.
The line equation for $\mcl_0$ \eqref{eq:line0} gives $y_2 = 1 - \frac{1}{qR_0} - y_1$. Then, in terms of $y_1$, we get
$y_1 (r_1 - r_2) < \frac{c_\mathcal{D}}{2} - r_2 \left(1 - \frac{1}{qR_0}\right)$.
The line equation also gives the minimum and maximum values for $y_1$ as $0$ and $1 - \frac{1}{qR_0}$, respectively.
Substituting into the inequality and solving yields the condition: \scriptsize
\begin{equation}\label{eq:cond:l0}
\left(1 - \frac{1}{qR_0} < \frac{c_\mathcal{D}}{2r_2}\right) \land
\left(1 - \frac{1}{qR_0} < \frac{c_\mathcal{D}}{2r_1}\right)
\end{equation}
\normalsize
By the theorem hypothesis, the condition in \eqref{eq:cond:l0} is satisfied. Therefore, $\lambda_3 < 0$. 
Thus, $\mcl_0$ is the center eigenspace of $J(\mcl_0)$, and the corresponding center manifold is stable \cite[Theorem~7.26]{sastry1999nonlinear}. 
If  $1 - \frac{1}{qR_0} > \frac{c_\mathcal{D}}{2r_1}$ or if $1-\frac{q}{R_0}>\frac{c_\mcd}{2r_2}$, then by a straightforward reversal of the arguments presented above, it can be seen that $\lambda_3>0$, which implies that $s(J(\mcl_0))>0$, which, from \cite[Theorem~5.42]{sastry1999nonlinear}, guarantees that the line $\mcl_0$ is unstable.~$\square$
\vspace{-2mm}
\subsection{Stability of line $\mathcal L_1$}
Our second main result is the following theorem.
\begin{theorem}\label{thm:l1}
    \seb{Consider system~\eqref{eq:main} under Assumptions~\ref{assume:betadelta} \ref{assume:q},\ref{assume:c}.
    Suppose further that $R_0 = \frac{\beta_1}{\delta_1} = \frac{\beta_2}{\delta_2}$ and $q^2 R_0 > 1$.
    The line $\mcl_1$ is locally exponentially stable if
    $1 - \frac{1}{q^2R_0} > \frac{c_\mathcal{D}}{2r_2}$ and $1 - \frac{1}{q^2R_0} > \frac{c_\mathcal{D}}{2r_1}$. The line $\mcl_1$ is unstable if $1 - \frac{1}{q^2R_0} < \frac{c_\mathcal{D}}{2r_2}$ or $1 - \frac{1}{q^2R_0} < \frac{c_\mathcal{D}}{2r_1}$.}
\end{theorem}
\textit{Proof:}
Under the hypothesis of the theorem, the conditions for Lemma~\ref{lem:coexist} statement (2) are satisfied; consequently, $\mcl_1$ is guaranteed to exist.
From Equation~\eqref{eq:line1}, we get $1 - y_1 - y_2 = s = \frac{1}{q^2 R_0}$; note the additional $q$ in the denominator of the RHS, when compared to the RHS of~\eqref{eq:line0}. Under Assumptions~\ref{assume:betadelta},\ref{assume:q}, $\frac{1}{q^2 R_0}\in(0,1)$.
The Jacobian for $\mcl_1$ can be put in terms of the Jacobian for $\mcl_0$: $J(\mcl_1) = \diag(q, q, -1, 1, 1) J(\mcl_0)$. Then, using an identical partitioning as in Theorem~\ref{thm:l0}, we get $J_1(\mcl_1) = q \cdot J_1(\mcl_0)$. Therefore, the eigenvalues of $J_1(\mcl_1)$ are the same as for $J_1(\mcl_0)$ scaled by $q$, and, since by Assumption~\ref{assume:q} $q$ is positive, do not differ in sign.
For $J_3(\mcl_1)$, we get the same eigenvalues as for $J_3(\mcl_0)$, except that the expression for $\lambda_3$ is negated.
This gives the reverse inequality to that in Theorem~\ref{thm:l0}; i.e.,  $2(r_1y_1 + r_2y_2) > c_\mcd$. By a similar derivation as in Theorem~\ref{thm:l0}, it is straightforward to show that if \scriptsize
\begin{equation}\label{eq:cond:l1:stable}
\left(1 - \frac{1}{q^2R_0} > \frac{c_\mathcal{D}}{2r_2}\right) \land
\left(1 - \frac{1}{q^2R_0} > \frac{c_\mathcal{D}}{2r_1}\right)
\end{equation}
\normalsize
then $\lambda_3 < 0$.
The rest of the  proof follows identically to the proof of stability of line $\mcl_0$ in Theorem~\ref{thm:l0}. 
Analogously, if \scriptsize
\begin{equation}\label{eq:cond:l1:unstable}
\left(1 - \frac{1}{q^2R_0} < \frac{c_\mathcal{D}}{2r_2}\right) \lor
\left(1 - \frac{1}{q^2R_0} < \frac{c_\mathcal{D}}{2r_1}\right),
\end{equation}
\normalsize
then it is clear that $\lambda_3 > 0$, which is sufficient for $\mcl_1$ to be unstable.~$\square$
\begin{remark}[Exclusivity of stability for $\mcl_0$ and $\mcl_1$]
Combining the stability conditions for $\mcl_0$ and $\mcl_1$ yields the inequality:
$1 - \frac{1}{qR_0} < \frac{c_\mathcal{D}}{2r_i} < 1 - \frac{1}{q^2R_0}, i\in[2]$, which yields the condition: $q R_0 < q^2R _0$. Under Assumptions~\ref{assume:betadelta} and \ref{assume:q} this is unsatisfiable. Therefore, $\mcl_0$ and $\mcl_1$ cannot simultaneously be stable.~$\square$
\end{remark}


\section{Simulations}\label{sec:sims}
For each type of equilibrium of system~\eqref{eq:main} identified in Section~\ref{sec:analysis},
we provide a brief description,
example parameters for its existence, and initial conditions
for which the dynamics of our system, where possible, converge to an equilibrium of interest; we provide several plots showing the system converging in simulation.

We select $x(0)$ in $\interior\Gamma$. Because the state variables often remain fixed at the boundary of their domain (e.g., $z_\S(t_k) = 1 \implies z_\S(t) = 1, \forall t \ge t_k$), we wish to demonstrate the system converging to the boundary equilibria, even when the system does not start on the boundary. 
Note that for stable equilibria we may always select initial conditions in $\interior\Gamma$ that converge to the equilibrium. Simulations were performed using fourth-order Runge-Kutta approximation with $h = 10^{-4}$.  

For ease of exposition, we write the indexed parameters in vector notation: $\beta := [\beta_1, \beta_2]^T$, $\delta := [\delta_1, \delta_2]^T$, $\bs{r} := [r_1, r_2]^T$, and $\bs{c} := [c_1, c_2]^T$.

\begin{figure}[h!]\centering
\includegraphics[width=0.45\textwidth]{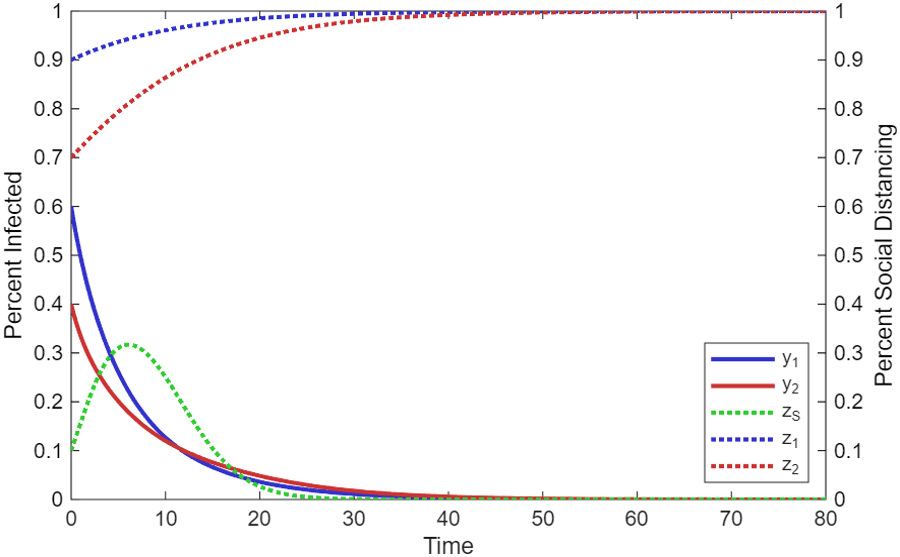}
\caption{Simulation showing both viruses dying out and the system converging to $\bs{p}_{\DFE 0}$.}
\label{figure:P_DFE0}
\end{figure}

Recall from Lemma~\ref{lem:DFE} that the fixed point $\bs{p}_{\DFE 0} := [0,0,0,1,1]^T$ always exists and is stable if $\delta_i > q\beta_i, i\in[2]$ (Prop.~\ref{prop:localstability:DFE0}); this is the standard condition $R_0 < 1$, mediated by the social distancing interaction factor $q$.
This gives two decoupled equations that are easily satisfied.
We choose $\beta=[\begin{smallmatrix}
    0.8&&0.8
\end{smallmatrix}]$, $\delta=[\begin{smallmatrix}
    0.2&&0.2
\end{smallmatrix}]$, $r=[\begin{smallmatrix}
    0.5&&0.5
\end{smallmatrix}]$, $c=[\begin{smallmatrix}
    0.5&&0.5
\end{smallmatrix}]$, $c_{\mcd}=0.4$ and $q=0.1$.
With this choice of parameters, the conditions for Proposition~\ref{prop:localstability:DFE0} are satisfied. 
Let $x(0) = [0.6, 0.4, 0.1, 0.9, 0.7]^T\in\interior\Gamma$. 
Observe that $y_1(0) + y_2(0) = 100\%$, and the fixed point $\bs{p}_{\DFE 0}$ is still stable and attractive, as demonstrated in Figure~\ref{figure:P_DFE0}.

\begin{figure}[h!]\centering
\includegraphics[width=0.45\textwidth]{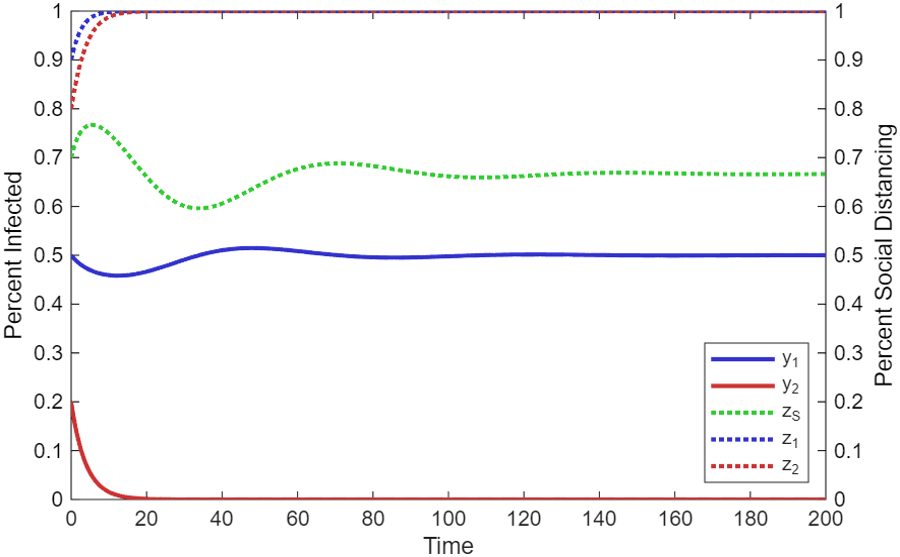}
\caption{Simulation showing virus~$1$ endemic and virus~$2$ dying out with partial social distancing in the healthy population, i.e., the system converging to $\bs{p}_{1\S}$.}
\label{figure:P_1S}
\end{figure}

From Lemma~\ref{lem:unilateral}, the existence of $\bs{p}_{1\S}$ requires $c_\mcd < 2 r_1$ so that $y_1\in(0,1)$ and $\beta_1(1 - \frac{c_\mcd}{2 r_1})q^2 < \delta_1 < \beta_1(1 - \frac{c_\mcd}{2 r_1})q$ so that $z_\S\in(0,1)$.
We choose $\beta=[\begin{smallmatrix}
    0.5&&0.4
\end{smallmatrix}]$, $\delta=[\begin{smallmatrix}
    0.06&&0.3
\end{smallmatrix}]$, $r=[\begin{smallmatrix}
    0.6&&0.4
\end{smallmatrix}]$, $c=[\begin{smallmatrix}
    1.0&&0.9
\end{smallmatrix}]$, $c_{\mcd}=0.6$ and $q=0.4$.
Observe that $\delta_1$ is relatively small in the example.
Let $x(0) = [0.5, 0.2, 0.7, 0.9, 0.8]^T\in\interior\Gamma$.
Then, we can observe, in Figure~\ref{figure:P_1S}, virus~$2$ dying out and virus~$1$ endemic, with partial social distancing in the healthy population.

\begin{figure}[h!]\centering
\begin{minipage}{0.25\textwidth}\centering
    \includegraphics[width=1\textwidth]{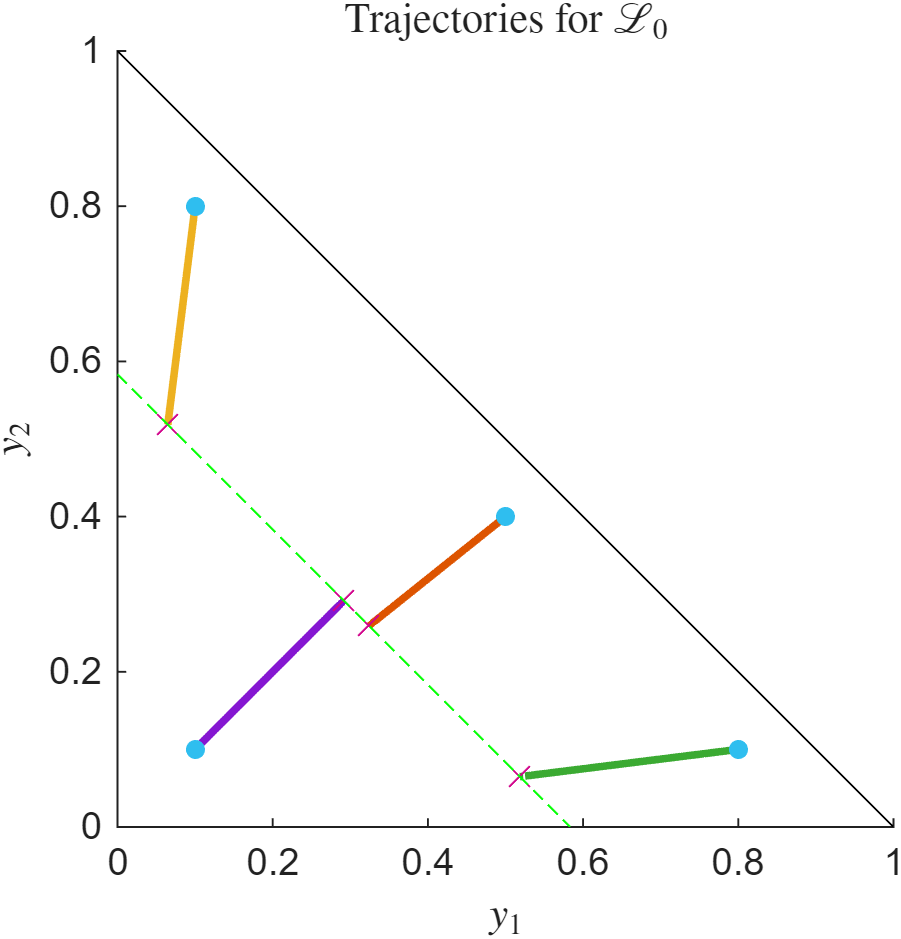}
    \caption{Simulation of $\mcl_{0}$.}
    \label{figure:L_0:traj}
\end{minipage}%
\begin{minipage}{0.25\textwidth}\centering
    \includegraphics[width=0.85\textwidth]{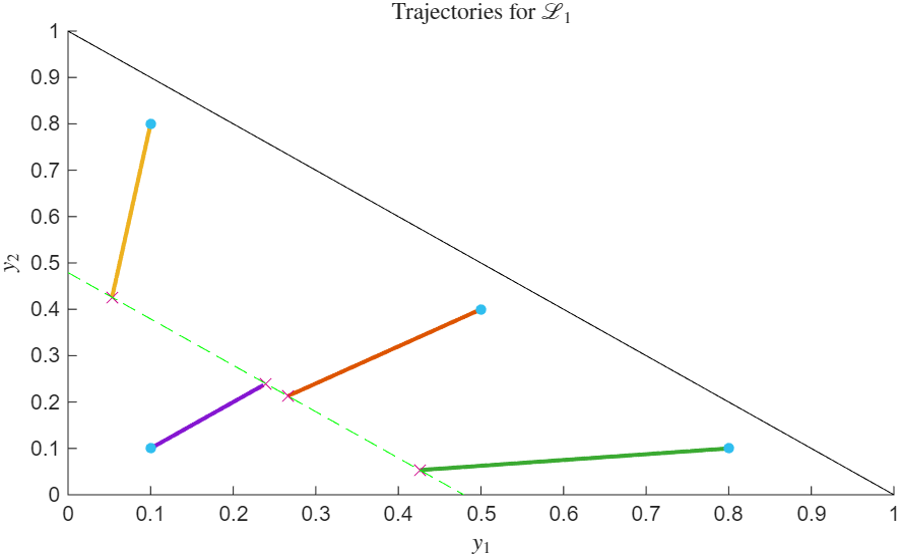}
    \caption{Simulation of $\mcl_{1}$.}
    \label{figure:L_1:traj}
\end{minipage}
\end{figure}

We choose
$\beta=[\begin{smallmatrix}
    0.3&&0.3
\end{smallmatrix}]$, $\delta=[\begin{smallmatrix}
    0.1&&0.1
\end{smallmatrix}]$, $r=[\begin{smallmatrix}
    0.5&&0.1
\end{smallmatrix}]$, $c=[\begin{smallmatrix}
    3&&3
\end{smallmatrix}]$, $c_{\mcd}=2$ and $q=0.8$.
Such a choice guarantees the existence of a line of coexistence equilibria, specifically the line $\mcl_0$ (see Equation~\eqref{eq:line0}). Moreover, said choice also satisfies the condition in Theorem~\ref{thm:l0} for $\mcl_0$ to be stable.
Here we show multiple trajectories converging to $\mcl_0$.
We set $z_\S = 0, z_1 = z_2 = 1$, so that we can project onto $\Delta$, \seb{where $\Delta$ is as defined in~\eqref{eq:Delta}}, without loss of information. We use the initial conditions $y(0)\in\{[0.5, 0.4], [0.1, 0.8], [0.1, 0.1], [0.8, 0.1]\}$, and plot the trajectories in Figure~\ref{figure:L_0:traj}. The initial conditions are shown as cyan dots; the equilibria points are shown as red crosses, and the colored lines represent the evolution of each initial condition. The line $\mcl_0$ is shown as the dashed green line.


The $\mcl_{1}$ case is nearly identical to the $\mcl_{0}$ case. The $qR_0$ term gains a $q$ becoming $q^2R_0 > 1$, and the stability condition is inverted to $2\bs{r}^T\bs{y} > c_\mcd$. Thus, we use similar parameters as for the $\mcl_0$, except we select $c_\mcd$ small enough s.t. $2 r_2 \left(1 - \frac{1}{qR_0}\right) > c_\mathcal{D}$.
We choose
$\beta=[\begin{smallmatrix}
    0.3&&0.3
\end{smallmatrix}]$, $\delta=[\begin{smallmatrix}
    0.1&&0.1
\end{smallmatrix}]$, $r=[\begin{smallmatrix}
    0.4&&0.4
\end{smallmatrix}]$, $c=[\begin{smallmatrix}
    0.9&&0.9
\end{smallmatrix}]$, $c_{\mcd}=0.1$ and $q=0.8$.
Using the same initial conditions as for the $\mcl_0$ example, we observe similar behavior in Figure~\ref{figure:L_1:traj}. Observe that the line $\mcl_1$ has a lower $y_2$-intercept than the example for $\mcl_0$, even though they use the same $\beta$, $\delta$, and $q$ parameters; this is because $q^2 R_0 < q R_0$.


\section{Conclusion}\label{sec:conclusions}
The paper proposed a model for the spread of two competing viruses in a single population, with the possibility of the susceptible individuals adopting (varying levels of) social distancing. The evolution of social distancing behaviors in the population is studied using replicator dynamics. Our main contributions were identification of the different kinds of equilibria that our model possesses. Moreover, we also secured conditions for (in)stability of $\bs{p}_{\DFE0}$, $\bs{p}_{\DFE1}$,$\bs{p}_{10}$ and $\bs{p}_{11}$. We also identified sufficient conditions for local asymptotic stability of lines of coexistence equilibria, namely $\mcl_0$ and $\mcl_1$. 
One line of future work could seek to understand what happens when the reproduction numbers of each virus~$1$ and virus~$2$ are greater than one- a case that is not covered in the present paper. Yet another thread of investigation could seek to design control strategies (possibly by using the perceived risk factor) for guaranteeing eradication of both the viruses. Finally, extending our model to the vector case would help better understand the effectiveness/limitations of our modeling framework.

\bibliographystyle{ieeetr}

\bibliography{keith}
\end{document}